\documentclass[twocolumn,showpacs,preprintnumbers,amsmath,amssymb]{revtex4}

\usepackage{graphicx}
\usepackage{dcolumn}
\usepackage{bm}

\begin{document}

\preprint{APS/123-QED}

\title{Variational principle of counting statistics in master equations}

\author{Jun Ohkubo} 
\email[Email address: ]{ohkubo@issp.u-tokyo.ac.jp}
\affiliation{
Institute for Solid State Physics, University of Tokyo, 
Kashiwanoha 5-1-5, Kashiwa-shi, Chiba 277-8581, Japan
}

\date{\today}

\begin{abstract}
We study counting statistics of number of transitions in a stochastic process.
For mesoscopic systems, a path integral formulation for the counting statistics has already been derived.
We here show that 
it is also possible to derive the similar path integral formulation 
without the assumption of mesoscopic systems.
It has been clarified that
the saddle point method for the path integral is not an approximation, but a valid procedure 
in the present derivation.
Hence, a variational principle in the counting statistics is naturally derived.
In order to obtain the variational principle, 
we employ many independent replicas of the same system.
In addition,
the Euler-Maclaurin formula is used in order to connect the discrete and continuous properties of the system.
\end{abstract}

\pacs{82.20.-w, 05.10.Gg, 04.20.Fy}
\maketitle

In a stochastic system,
it would be possible to count a number of a certain target transition,
and the number of the target transition is a random variable.
Since the random variable is directly related to a net current or flow in the system,
the counting of number of transitions plays an important role in nonequilibrium physics,
Especially, 
nonequilibrium properties for time-dependent systems have been studied largely;
e.g., Brownian motors (ratchet systems) \cite{Julicher1997,Reimann2002,Astumian2002}
and a pump current problem
\cite{Liu1990,Tsong2003,Astumian2003,Sinitsyn2007,Sinitsyn2007a,Ohkubo2008,Ohkubo2008a,Sinitsyn2008}. 

There are several methods to calculate average current 
in periodically time-dependent systems
\cite{Astumian1989,Robertson1991,Jain2007}.
The calculation of the average current enables us to have
the no-pumping theorem and the pumping restriction theorem,
which will become useful for the study of biochemical reactions or molecular machines
\cite{Rahav2008,Chernyak2008}.
In addition, recent progress gives recipes to evaluate not only the average current,
but also all statistics including the fluctuation, i.e., `counting statistics'
\cite{Sinitsyn2007,Sinitsyn2007a,Ohkubo2008,Ohkubo2008a,Sinitsyn2008}.
We note that 
basic concepts of full counting statistics (FCS) in condensed matter physics 
\cite{Pilgram2003,Jordan2004,Bagrets2006}
are similar to the classical stochastic cases.

For the counting statistics in master equations,
there are mainly two different approaches.
One is the transition matrix approach \cite{Gopich2003,Gopich2005,Gopich2006},
and the extensions of this approach have succeeded in calculating the pump current analytically
\cite{Sinitsyn2007,Ohkubo2008,Ohkubo2008a}.
The other one is based on a path integral formulation \cite{Sinitsyn2007a}.
The transition matrix approach requires diagonalization of an effective evolution Hamiltonian,
which may be a complicated task for a system with a large transition matrix such as the ratchet problem.
In contrast, the path integral formulation is available for the ratchet problem \cite{Sinitsyn2007a}.

In \cite{Sinitsyn2007a}, 
a path integral formulation for the counting statistics in mesoscopic systems has been investigated.
The word `mesoscopic' means that the system contains many particles or elements,
so that the saddle point method in the analysis is approximately justified.
Hence, one can finally obtain saddle point equations similar to the Hamilton's canonical equations 
in classical mechanics.
In the FCS, the similar discussions have been used \cite{Pilgram2003,Jordan2004}.
For the mesoscopic systems,
a replacement of a discrete variable with a continuous variable 
(or replacement of a summation with an integral)
may be justified because the discrete variable takes a large value.
However, if one loses the assumption of mesoscopic systems,
the validity of the usage of the saddle point method 
and the replacement of the discrete variable with a continuous one are ambiguous.

In the present paper, we will clarify that
the saddle point equations similar to the Hamilton's canonical equations 
are derived without the assumption of mesoscopic systems.
Furthermore, it is clarified that
the saddle point method is not an approximation,
but a valid procedure in our derivation.
Hence, we can finally obtain a kind of variational principle in the counting statistics.
In order to obtain it,
many replicas of microscopic systems are adopted,
and the Euler-Maclaurin formula is applied
in order to replace a discrete variable with a continuous one.
This variational principle would become a basis for the counting problem,
because it will allow us to use many analytical techniques developed in the Hamilton systems
in order to investigate the counting statistics in stochastic systems.
To demonstrate the derivation,
we consider a simple model with $2 \times 2$ transition matrix in the present paper.
Complicated models such as a ratchet problem could also be treated in a similar manner,
by using a similar discussion in \cite{Sinitsyn2007a}.

Let us consider a simple model introduced in \cite{Sinitsyn2007} for a pump current problem.
In \cite{Sinitsyn2007a}, the similar system in the mesoscopic regime (the number of particles is very large)
has been studied, but we here investigate the original system in \cite{Sinitsyn2007},
which does not have the mesoscopic property.
We here treat the following model:
\begin{align}
[\mathrm{L}] \leftrightarrows [\mathrm{bin}] \leftrightarrows [\mathrm{R}],
\end{align}
where $[\mathrm{L}]$ and $[\mathrm{R}]$ are particle baths,
and $[\mathrm{bin}]$ is assumed to contain either zero or one particles in it.
The full kinetic scheme is

(i) $[\mathrm{L}] \rightarrow [\mathrm{bin}]$ with rate $k_1(t)$;

(ii) $[\mathrm{L}] \leftarrow [\mathrm{bin}]$ with rate $k_{-1}(t)$;

(iii) $[\mathrm{bin}] \rightarrow [\mathrm{R}]$ with rate $k_2(t)$;

(iv) $[\mathrm{bin}] \leftarrow [\mathrm{R}]$with rate $k_{-2}(t)$. 

\noindent
The transition rates $\{k_i(t)\}$ can be time-dependent,
and in what follows, we denote them as $\{k_i\}$ for simplicity.
The master equation is written as follows:
\begin{align}
\frac{d}{d t}
\begin{bmatrix}
p \\ 1-p
\end{bmatrix}
= 
\begin{bmatrix}
-k_1 - k_{-2} & k_{-1} + k_2 \\
k_1 + k_{-2} & -k_{-1} - k_2
\end{bmatrix}
\begin{bmatrix}
p \\ 1-p
\end{bmatrix},
\end{align}
where $p$ is the probability that $[\mathrm{bin}]$ is empty.
Our goal is to evaluate the $[\mathrm{bin}] \to [\mathrm{R}]$ flux for a finite time $T$.

In order to calculate the counting statistics, we use the path integral technique.
At first, we discretize the time as $t = m \delta t$, $m \in \mathbf{N}$.
Hence, the final time is characterized by an integer $M$ via $T = M \delta t$.
The probability $p$ is time-dependent, so that we denote the probability $p$ at time $t=m \delta t$
as $p_m$ in order to show the time dependence of $p$ explicitly.

We here define the number of reaction (or hopping) $i \in \{1,-1,2,-2\}$ 
at time step $m$ as $\Delta \tilde{Q}_m^{(i)}$.
A stochastic process described by a master equation can be simulated exactly
by the Gillespie algorithm \cite{Gillespie1977,Anderson2007},
and the number of reaction $i$ during time $\delta t$ obeys the Poissonian
with average $k_i p_{m-1} \delta t$.
For example, the probability of $\Delta \tilde{Q}_m^{(1)}$  may be given by
\begin{align}
P(\Delta \tilde{Q}_m^{(1)} | p_{m-1})
= \exp \left(-k_1 p_{m-1} \delta t \right) 
\frac{(k_1 p_{m-1}  \delta t)^{\Delta \tilde{Q}_m^{(1)}}}{(\Delta \tilde{Q}_m^{(1)})!}.
\label{eq_probability_hopping}
\end{align}
An occurrence of a reaction changes the state,
but the state variable $p$ is continuous and $\Delta \tilde{Q}_m^{(i)}$ is discrete.
In order to connect these quantities,
we discretize the continuous variable $p_m \in [0,1]$ as $p_m = n_m \delta n$,
where $n_m \in \{0, 1, 2, \dots, N_\mathrm{max}\}$ and $\delta n \equiv 1 / N_\mathrm{max}$.
Note that we can recover the continuous property of the probability $p_m$
when we take $\delta n \to 0$ (i.e., $N_\mathrm{max} \to \infty$).
This is the most important point in the present paper;
we should finally take $N_\mathrm{max} \to \infty$ 
in order to recover the continuous property of $p_m$.

There is still one problem as follows.
In the Gillespie algorithm, 
one reaction changes the state of the system, and then
the next reaction should be evaluated by using the updated state.
Hence, only one reaction step, i.e., $\Delta \tilde{Q}_m^{(i)} = 0$ or $1$, should be allowed.
However, the Poisson statistics (eq.~\eqref{eq_probability_hopping}) does not include such effects.
In other words, we should restrict $\Delta \tilde{Q}_m^{(i)}$ to $\{0,1\}$
for the natural interpretation of the Gillespie algorithm,
but the domain of \eqref{eq_probability_hopping} is $\{0,1,\dots\} = \mathbf{N}$;
it seems inadequate to use \eqref{eq_probability_hopping} directly.

In order to avoid this problem,
we introduce many independent replicas of the system with the same state $p_{m-1}$ (or $n_{m-1}$)
at each time step;
totally $N_\mathrm{max}$ systems with the same initial state $n_{m-1}$.
The total number of reaction $i$ in the $N_\mathrm{max}$ systems is therefore
\begin{align}
\Delta Q_m^{(i)} = \sum_{j=1}^{N_\mathrm{max}} \Delta \tilde{Q}_m^{(i)},
\end{align}
and the probability of $\Delta Q_m^{(1)}$ is given by
\begin{align}
P(\Delta Q_m^{(1)} | n_{m-1}) 
= \exp \left(-k_1 n_{m-1} \delta t \right) \frac{(k_1 n_{m-1} \delta t)^{\Delta Q_m^{(1)}}}{(\Delta Q_m^{(1)})!}.
\label{eq_probability_hopping_improved}
\end{align}
Here, we used the independent property of each replica.
$\{\Delta Q_m^{(i)}\}$ for $i=-1,2,-2$ are obtained by the similar manner.
Note that $\Delta Q_m^{(1)}$ is the \textit{total} number of reaction $1$ in $N_\mathrm{max}$ replicas,
and therefore $\Delta Q_m^{(1)} \in \mathbf{N}$ (not $\{0,1\}$).
If we choose $\delta t$ small enough,
it would be possible to assume that there is at most one reaction in each replica 
(after taking $N_\mathrm{max} \to \infty$).
The introduction of the independent replicas, therefore, enables us
to avoid the above problem (i.e., $\Delta \tilde{Q}_m^{(i)}$ should be $0$ or $1$).
In addition, we will see later that 
the introduction of the replicas is necessary to use the saddle point method.

In order to count the flux $[\mathrm{bin}] \to [\mathrm{R}]$,
we introduce the quantities 
$\tilde{Q}_m^\mathrm{R} = \Delta \tilde{Q}_m^{(2)} - \Delta \tilde{Q}_m^{(-2)}$ and
$Q_m^\mathrm{R} = \Delta Q_m^{(2)} - \Delta Q_m^{(-2)}$,
and calculate the characteristic function of $Q_m^\mathrm{R}$ during the time interval $T$.
The characteristic function conditioned by the initial state $n_0$
is calculated from
\begin{align}
&\mathbf{E} [ e^{i \chi_\mathrm{C} (Q_1^{\mathrm{R}} + Q_2^{\mathrm{R}} 
+ \cdots + Q_{M-1}^{\mathrm{R}}  + Q_M^{\mathrm{R}} )} 
| n_0] \nonumber \\
&= \mathbf{E} [ e^{i \chi_\mathrm{C} Q_1^{\mathrm{R}}} 
\mathbf{E} [ e^{i (\chi_\mathrm{C} Q_2^{\mathrm{R}} + \cdots + \chi_\mathrm{C} Q_M^{\mathrm{R}})}  | n_{1}]
| n_{0}] \nonumber \\ 
&= \cdots \nonumber \\
&= \mathbf{E} [ e^{i \chi_\mathrm{C} Q_1^{\mathrm{R}}} \mathbf{E} [ 
e^{i \chi_\mathrm{C} Q_2^{\mathrm{R}}} \mathbf{E} [ 
\cdots
\mathbf{E} [ e^{i \chi_\mathrm{C} Q_M^{\mathrm{R}}} | n_{M-1} ]
| \cdots ] | n_{0}],
\label{eq_def_characteristic_function}
\end{align}
where $\chi_\mathrm{C}$ may be assumed to be real.
In general, a characteristic function is rewritten in a form of $\exp[S(\chi_\mathrm{C})]$
(see \eqref{eq_chara_1} and \eqref{eq_chara_2}).
When we define a quantity $Q^\textrm{R} \equiv \sum_{m=1}^M Q_m^\textrm{R}$,
derivatives of $S(\chi_\mathrm{C})$ give cumulants for $Q^\mathrm{R}$;
e.g., 
$\langle Q^\mathrm{R} \rangle 
= - i \partial S(\chi_\mathrm{C}) / \partial \chi_\mathrm{C} |_{\chi_\mathrm{C} = 0}$, 
$\langle \delta^2 Q^\mathrm{R} \rangle 
\equiv \langle (Q^\mathrm{R})^2 \rangle - \langle Q^\mathrm{R} \rangle^2
= (-i)^2 \partial^2 S(\chi_\mathrm{C}) / \partial \chi^2 |_{\chi_\mathrm{C}=0}$,
etc.
The quantity $Q^\mathrm{R}$ corresponds to the flux [bin] $\to$ [R] 
in the $N_\textrm{max}$ systems,
and hence an explicit calculation for the characteristic function 
\eqref{eq_def_characteristic_function} is needed
to obtain the counting statistics for the flux [bin] $\to$ [R].

For simplicity, we here show a calculation for 
$\mathbf{E}[e^{i \chi_\mathrm{C} Q_m^{\mathrm{R}}} | n_{m-1} ]$;
the calculation of the characteristic function \eqref{eq_def_characteristic_function}
can be performed in a similar manner.
By using the Fourier transformation of the Kronecker delta,
\begin{align}
\delta(A,B) = \int_{-\pi}^{\pi} \frac{d\chi}{2\pi} e^{i \chi (A-B)}
\end{align}
we obtain
\begin{align}
&\mathbf{E}\left[e^{i \chi_\mathrm{C} Q_m^{\mathrm{R}}} 
| n_{m-1} \right] \nonumber \\
&= \frac{1}{Z_m} \sum_{n_m=0}^{N_\mathrm{max}} 
\sum_{\Delta Q_m^{(1)} = 0}^{\infty} \sum_{\Delta Q_m^{(-1)} = 0}^{\infty}
\sum_{\Delta Q_m^{(2)} = 0}^{\infty} \sum_{\Delta Q_m^{(-2)} = 0}^{\infty} \nonumber \\
&\Big\{ P(\Delta Q_m^{(1)}|n_{m-1}) P(\Delta Q_m^{(-1)}|n_{m-1}) \nonumber \\
&\times P(\Delta Q_m^{(2)}|n_{m-1}) P(\Delta Q_m^{(-2)}|n_{m-1}) 
\exp\left[ i \chi_\mathrm{C} Q_m^{\mathrm{R}} \right] \nonumber \\
&\times \delta\left( n_m - n_{m-1}, 
- \Delta Q_m^{(1)} - \Delta Q_m^{(-2)} + \Delta Q_m^{(-1)} + \Delta Q_m^{(2)} \right)
\Big\}
\nonumber \\
&= \frac{1}{Z_m} \sum_{n_m=0}^{N_\mathrm{max}} \int_{-\pi}^{\pi} \frac{d \chi_m}{2\pi}
\Big\{ e^{i \chi_m (n_m - n_{m-1})} 
\exp\left[ k_1 n_{m-1}  e_{+\chi_m} \delta t\right] \nonumber \\
&\times \exp\left[ k_2 (N_\mathrm{max} - n_{m-1})  e_{-\chi_m + \chi_\mathrm{C}} \delta t\right] \nonumber \\
&\times \exp\left[ k_{-1} (N_\mathrm{max} - n_{m-1})  e_{-\chi_m} \delta t\right] \nonumber \\
&\times \exp\left[ k_{-2} n_{m-1}  e_{+\chi_m-\chi_\mathrm{C}} \delta t\right]
\Big\},
\label{eq_one_step_path_integral}
\end{align}
where $e_{\pm \chi} \equiv e^{\pm i \chi} -1$.
Note that the probabilities $P(\Delta Q_m^{(i)}|n_{m-1}) $ for $i=\pm 1, \pm 2$
are correlated via the Kronecker delta,
and then a normalization constant $Z_m$ is introduced.

Equation~\eqref{eq_one_step_path_integral} has still one summation for $n_m$, 
and its discrete property is inconvenient for further analytical treatments.
However, different from the mesoscopic case in \cite{Sinitsyn2007a},
it may be inadequate in the present case to simply replace the summation for $n_m$ with an integral.
We therefore use the Euler-Maclaurin formula for a function $g(x)$:
\begin{align}
&\frac{1}{L} \sum_{n=n_1}^{n_2} g\left( \frac{n}{L} \right) \nonumber \\
&= \int_{(n_1-1/2)/L}^{(n_2+1/2)/L} g(x) d x \nonumber \\
&- \frac{1}{24L^2} 
\left[ g'\left(\frac{n_2 + 1/2}{L}\right) 
- g'\left(\frac{n_1 - 1/2}{L}\right)
\right] + \cdots,
\end{align}
where $g'(x)$ is the derivative of $g(x)$ with $x$.
In our calculation, we have the following form with a function $f(n)$:
\begin{align}
&\sum_{n=0}^{N_\mathrm{max}} e^{N_\mathrm{max} f(n/N_\mathrm{max})} \nonumber \\
&= N_\mathrm{max} \int_{-1/(2N_\mathrm{max})}^{1 + 1/(2N_\mathrm{max})} e^{N_\mathrm{max} f(x)} d x \nonumber \\
&- \frac{1}{24} 
\left[ f'\left(1+\frac{1}{2 N_\mathrm{max}} \right) e^{N_\mathrm{max} f(1+\frac{1}{2 N_\mathrm{max}})} \right.
\nonumber \\
&\left.
- f'\left(-\frac{1}{2 N_\mathrm{max}} \right) e^{N_\mathrm{max} f(-\frac{1}{2 N_\mathrm{max}})}
\right] + \dots,
\label{eq_eular_maclaurin}
\end{align}
where $f(x)$ is determined by \eqref{eq_one_step_path_integral}
(or eventually, by \eqref{eq_def_characteristic_function}).
Although $f(x)$ is a complex function,
we can easily see that the real part of $f(x)$ is negative because $\mathrm{Re}\, e_{\pm \chi} \le 0$
in \eqref{eq_one_step_path_integral}.
Hence, the second term in \eqref{eq_eular_maclaurin} vanishes when $N_\mathrm{max} \to \infty$,
and we can verify the replacement of the discrete variable with the continuous variable.
Note that the domain of the integration becomes $[0,1]$ when $N_\mathrm{max} \to \infty$;
this is consistent with the fact that $p$ is the probability.

Finally, we obtain the characteristic function \eqref{eq_def_characteristic_function}
in the path integral form as
\begin{align}
&\mathbf{E} [ e^{i \chi_\mathrm{C} (Q_1^{\mathrm{R}} + Q_2^{\mathrm{R}} 
+ \cdots + Q_{M-1}^{\mathrm{R}}  + Q_M^{\mathrm{R}} )} 
| p_0] \nonumber \\
&\sim \prod_{m=1}^M \left[
\int_{- \frac{1}{2N_\mathrm{max}}}^{1+\frac{1}{2N_\mathrm{max}}} d p_m \int_{-\pi}^{\pi} \frac{d \chi_m}{2\pi} \right]
 e^{N_\mathrm{max} S(\chi_\mathrm{C},\{p_m\},\{\chi_m\})},
\end{align}
where
\begin{align}
&S(\chi_C, \{p_m\},\{\chi_m\}) \nonumber \\
&= \sum_{m=1}^M
\delta t \left[ 
i \chi_m \frac{p_m - p_{m-1}}{\delta t}
- H(\chi_\mathrm{C},p_{m-1},\chi_m)
\right],
\end{align}
and
\begin{align}
H & (\chi_\mathrm{C},p_{m-1},\chi_m) \nonumber \\
=& -k_1 p_{m-1} e_{+\chi_m} - k_2 (1-p_{m-1}) e_{-\chi_m+\chi_\mathrm{C}} \nonumber \\
& - k_{-1} (1-p_{m-1}) e_{-\chi_m} - k_{-2} p_{m-1} e_{+\chi_m-\chi_\mathrm{C}}.
\end{align}

Note that $N_\mathrm{max} \to \infty$ is a necessary condition
in order to recover the continuous property of $\{p_m\}$.
Hence, we use the saddle point method and obtain 
\begin{align}
\mathbf{E} [ e^{i \chi_\mathrm{C} \sum_m Q_m^{\mathrm{R}}} | p_0] 
\sim \exp \left[ N_\mathrm{max} S(\chi_\mathrm{C}, \{p_m\}^{\mathrm{cl}}, \{\chi_m\}^{\mathrm{cl}})
\right].
\label{eq_chara_1}
\end{align}
In addition, 
the random variables $\{\tilde{Q}_m^\mathrm{R}\}$ are independent each other,
so that 
\begin{align}
\mathbf{E} [ e^{i \chi_\mathrm{C} \sum_m \tilde{Q}_m^{\mathrm{R}}} | p_0] 
\sim \exp \left[ S(\chi_\mathrm{C}, \{p_m\}^{\mathrm{cl}}, \{\chi_m\}^{\mathrm{cl}})
\right],
\label{eq_chara_2}
\end{align}
where the superscript `cl' means that
these quantities are determined by saddle point equations
(see \eqref{eq_motion_1} and \eqref{eq_motion_2}).
After taking the continuous time limit,
we finally obtain the following result
\begin{align}
&\mathbf{E} [ e^{i \chi_\mathrm{C} \int_0^T d t\tilde{Q}^{\mathrm{R}}(t)} | p_0] \nonumber \\
&\sim \exp\left[ \int_0^T d t \left( 
i \chi^\mathrm{cl}(t) \frac{d p^\mathrm{cl}(t)}{d t}
- H(\chi_\mathrm{C}, p^\mathrm{cl}(t), \chi^\mathrm{cl}(t))
\right)\right],
\label{eq_final_result}
\end{align}
where $\chi^\mathrm{cl}(t)$ and $p^\mathrm{cl}(t)$
are evaluated by the saddle point equation
\begin{align}
&i \frac{d p^\mathrm{cl}(t)}{d t}
= + \frac{\partial H(\chi_\mathrm{C},p^\mathrm{cl}(t),\chi^\mathrm{cl}(t))}{\partial \chi^\mathrm{cl}(t)}.
\label{eq_motion_1} \\
&i \frac{d \chi^\mathrm{cl}(t)}{d t}
= - \frac{\partial H(\chi_\mathrm{C},p^\mathrm{cl}(t),\chi^\mathrm{cl}(t))}{\partial p^\mathrm{cl}(t)},
\label{eq_motion_2}
\end{align}
Different from the mesoscopic case in \cite{Sinitsyn2007a},
the saddle point method is not approximation, as explained above;
\eqref{eq_final_result}, \eqref{eq_motion_1} and \eqref{eq_motion_2} are valid because
we finally take $N_\mathrm{max} \to \infty$.
As a result, we obtain a kind of \textit{variational principle} in the counting statistics.
Namely, when we define the \textit{action} $S$ as
\begin{align}
S = \int \left( i \chi \frac{d p}{dt} - H \right) dt,
\end{align}
the path of $\chi$ and $p$ are given by the variational principle,
and then we have equations similar to Hamilton's canonical equations \eqref{eq_motion_1} and 
\eqref{eq_motion_2}.

Using the variational principle, we can calculate a pump current when
the transition rates $k_1$ and $k_{-2}$ vary adiabatically with a period $T_0$
($k_{-1}$ and $k_2$ are assumed to be time-independent).
Following the similar analytical treatment in \cite{Sinitsyn2007a},
we define $\mathbf{S_c}$ as the surface enclosed by the contour $\mathbf{c}$ 
in the space of parameters $k_1$ and $k_{-2}$,
and set the time derivatives in \eqref{eq_motion_1} and \eqref{eq_motion_2} as zero
due to the adiabatic condition.
We finally obtain
\begin{align}
\mathbf{E} [ e^{i \chi_\mathrm{C} \int_0^T d t\tilde{Q}^{\mathrm{R}}(t)} | p_0] \sim e^{S_\mathrm{geom} + S_\mathrm{cl}},
\end{align}
where 
\begin{align}
S_\mathrm{geom} = \frac{T}{T_0} 
\oint_\mathbf{S_c} dk_1 dk_{-2} F_{k_1,k_{-2}},
\end{align}
\begin{align}
F_{k_1,k_{-2}} = - e_{-\chi_\mathrm{C}} \left(e^{i \chi_\mathrm{C}} k_2 + k_{-1} \right) K^{-3},
\end{align}
and
\begin{align}
S_\mathrm{cl} = \frac{-T}{2T_0} 
\int_0^{T_0} dt (K_+ - K).
\end{align}
Here, we define $K_+ \equiv k_1 + k_{-2} + k_{-1} + k_2$
and $K \equiv (k_+^2 4 k_1 k_2 e_{\chi_\mathrm{C}} + 4 k_{-1} k_{-2} e_{-\chi_\mathrm{C}})^{1/2}$.
This is consistent with the results in \cite{Sinitsyn2007}.

In conclusion, we derived a variational principle for a counting statistics
in a microscopic master equation.
The word `microscopic' means that we do not need an assumption of the mesoscopic systems.
Due to the lack of the mesoscopic properties,
we cannot simply connect discrete variables and continuous variables.
Hence, the derivation of the path integral formulation was carefully performed.
In the derivation, the continuous state variable is replaced with a discrete variable,
and the condition to recover the continuous property
validates the usage of the saddle point method.
One of advantages of this variational scheme is that
we may use many analytical method in Hamilton systems
in order to study the counting statistics in stochastic processes.
While further studies will be needed
in order to make the variational principle given in the present paper useful,
the present work will give a basis for future works of counting statistics.

We thank M. Oshikawa for helpful discussions,
and N. A. Sinitsyn for helpful comments for this work.
This work was supported in part by grant-in-aid for scientific research 
(Nos. 20115009 and 21740283)
from the Ministry of Education, Culture, Sports, Science and Technology, Japan.

\end{document}